\documentclass[aps,prl,10pt,twocolumn,superscriptaddress]{revtex4-1}

\usepackage{graphicx}
\usepackage{amsmath}
\usepackage{amsfonts}
\usepackage{amssymb}
\usepackage{color}
\usepackage{hyperref}
\hypersetup{colorlinks=true,allcolors=blue}

\usepackage{textcomp}

\begin{document}
\title{Phonon-induced enhancement of photon entanglement in quantum dot-cavity systems}
\author{T. Seidelmann}
\affiliation{Universit{\"a}t Bayreuth, Lehrstuhl f{\"u}r Theoretische Physik III, Universit{\"a}tsstra{\ss}e 30, 95447 Bayreuth, Germany}
\author{F. Ungar}
\affiliation{Universit{\"a}t Bayreuth, Lehrstuhl f{\"u}r Theoretische Physik III, Universit{\"a}tsstra{\ss}e 30, 95447 Bayreuth, Germany}
\author{A. M. Barth}
\affiliation{Universit{\"a}t Bayreuth, Lehrstuhl f{\"u}r Theoretische Physik III, Universit{\"a}tsstra{\ss}e 30, 95447 Bayreuth, Germany}
\author{A. Vagov}
\affiliation{Universit{\"a}t Bayreuth, Lehrstuhl f{\"u}r Theoretische Physik III, Universit{\"a}tsstra{\ss}e 30, 95447 Bayreuth, Germany}
\affiliation{ITMO University, St. Petersburg, 197101, Russia}
\author{V. M. Axt}
\affiliation{Universit{\"a}t Bayreuth, Lehrstuhl f{\"u}r Theoretische Physik III, Universit{\"a}tsstra{\ss}e 30, 95447 Bayreuth, Germany}
\author{M. Cygorek}
\affiliation{Department of Physics, University of Ottawa, Ottawa, Ontario, Canada K1N 6N5}
\author{T. Kuhn}
\affiliation{Institut f{\"u}r Festk{\"o}rpertheorie, Universit{\"a}t M{\"u}nster, 48149 M{\"u}nster, Germany}

\begin{abstract}
We report on simulations of the degree of polarization entanglement of photon pairs simultaneously emitted from a quantum dot-cavity system that demand revisiting the role of phonons. Since coherence is a fundamental precondition for entanglement and phonons are known to be a major source of decoherence, it seems unavoidable that phonons can only degrade entanglement. In contrast, we demonstrate that phonons can cause a degree of entanglement that even surpasses the corresponding value for the  phonon-free case. In particular, we consider the situation of comparatively small biexciton binding energies and either finite exciton or cavity mode splitting. In both cases, combinations of the splitting and the dot-cavity coupling strength are found where the entanglement exhibits a nonmonotonic temperature dependence which enables entanglement above the phonon-free level in a finite parameter range. This unusual behavior can be explained by phonon-induced renormalizations of the dot-cavity coupling $g$ in combination with a nonmonotonic dependence of the entanglement on $g$ that is present already without phonons.
\end{abstract}
\maketitle

The appearance of entangled states is one of the showcase effects that highlights most impressively the dramatic conceptual changes brought forth by going over from classical to quantum physics \cite{Horodecki:09,audretsch:07}.
Moreover, realizations of entangled states, mostly with photons, have paved the way toward many innovative applications \cite{Zeilinger_entangled}, e.g., in quantum cryptography \cite{stevenson:02,Gisin:02}, quantum teleportation \cite{Bouwmeester:97}, quantum information processing \cite{pan:12,Bennett:00,audretsch:06,Kuhn:16}, and photonics \cite{Photonic_quantum_technologies}.
In particular, quantum dot (QD) cavity systems have attracted a lot of attention  as sources for triggered
entangled photon pairs \cite{Stevenson2006,Hafenbrak,dousse:10,EdV,entangled-photon1,entangled-photon2,Orieux_entangled,munoz15}, not only because these systems hold the promise of a natural integration in solid-state devices.
Embedding a QD in a microcavity enables the manipulation of few-electron and few-photon states in a system with high optical nonlinearities, which can be used for realizing a few-photon logic in quantum optical networks \cite{faraon11}.
Furthermore, the cavity boosts the quantum yield due to the Purcell effect \cite{Badolato:05,dousse:10} and, for high cavity quality factors Q, it reduces the detrimental effects of phonons on the photon indistinguishability \cite{grange17}.

The essence of entanglement in a bipartite system is the creation of a state that cannot be factorized into parts referring to the constituent subsystems, which requires the buildup of a superposition state.
Polarization entanglement between horizontally ($H$) or vertically ($V$) polarized photon pairs is established, e.g., by creating superpositions of the states $|HH\rangle$ and $|VV\rangle$ with two photons with either $H$ or $V$ polarizations exploiting the biexciton cascade \cite{Stevenson2006,Hafenbrak,dousse:10,EdV,entangled-photon1,entangled-photon2,Orieux_entangled}.
Starting from the biexciton, the system can decay  first into one of the two excitons and a photon with the corresponding polarization ($H$ or $V$).
The excitons then decay further to the QD ground state emitting a second photon with the same polarization as in the biexciton decay.
Ideally, the resulting quantum state is a coherent superposition and maximally entangled.
Which-path information introduced, e.g., by the fine-structure splitting of the excitons, leads to an asymmetric superposition and decreased entanglement.
The system can also decay from the biexciton directly to the ground state by simultaneous two-photon emission, a process which is much less affected by which-path information than the sequential single-photon decay \cite{Jahnke2012,heinze17,Seidlemann2019}.

Obviously, maintaining a coherent superposition requires stable relative phases between the involved states.  However, in a solid-state system, the interaction with the environment unavoidably leads to a loss of phase
coherence.
In particular, phonons are known to provide a major source of decoherence \cite{Knorr2003,inideco,dot_biex,machnikowski:08,ramsay:10,Hughes_QDcavity_strongdriving_2011,close2012,Kaer14,Doris2017,PhysRevLett.118.233602}, which led to the expectation that phonons should always degrade the entanglement.
Indeed, recent simulations \cite{Harouni:2014,heinze17,Different-Concurrences:18} are in line with this expectation.

In this Letter, we demonstrate that the phonon influence is not necessarily destructive.
On the contrary, phonons can increase the degree of photon entanglement when the destructive effect resulting from phonon-induced decoherence is overcompensated by phonon-related renormalizations of the QD-cavity coupling that shift the system into a regime of higher photon entanglement.
A precondition of this mechanism  is a decrease of the degree of entanglement with rising  QD-cavity coupling $g$ in the phonon-free case in a finite $g$ range.
This is realized, e.g., in the limit of weak biexciton binding and finite exciton or cavity mode splitting.
In both cases, the phonon-induced enhancement is found in a finite range of binding energies and couplings $g$. 

Our studies are based on the  Hamiltonian \cite{heinze17,Different-Concurrences:18}: 
\begin{align}
 \hat{H}=&\,\hbar\omega_H\vert X_H\rangle\langle X_H\vert+\hbar\omega_V\vert X_V\rangle\langle X_V\vert
                   \notag\\&
+\hbar(\omega_{H}+\omega_{V}-\omega_{B}) \vert B\rangle\langle B\vert
  +\sum_{\ell=H,V}\hbar\omega_\ell^c\,\hat{a}_\ell^\dagger\hat{a}_\ell
\notag\\&
+\sum_{\bf q}\hbar\omega_{\bf q}\hat{b}_{\bf q}^\dagger\hat{b}_{\bf q}
+\sum_{{\bf q},\chi}n_\chi\left(\gamma_{\bf q}
\hat{b}_{\bf q}^\dagger+\gamma_{\bf q}^*\hat{b}_{\bf q}\right)\vert \chi\rangle\langle \chi\vert
+ \hat{\mathcal{X}},
\end{align}
where $\vert B\rangle$ is the biexciton state with energy $\hbar(\omega_{H}+\omega_{V}-\omega_{B})$ and a biexciton binding energy $E_{B}=\hbar\omega_{B}$, while $\vert X_{H/V}\rangle$ denote the two exciton states with energies $\hbar\omega_{H/V}$ that couple to $H$ or $V$ polarized cavity modes with destruction (creation) operators $\hat{a}_{H/V}\,(\hat{a}^{\dag}_{H/V})$ and mode energies $\hbar\omega_{H/V}^c$. 
$\hat{b}_{\bf q}\,(\hat{b}^{\dag}_{\bf q})$ are operators that destroy (create) longitudinal acoustic phonons with wave vector ${\bf q}$ and energy $\hbar\omega_{\bf q}$.
We consider bulk phonons with a linear dispersion and account for the deformation potential coupling $\gamma_{\bf q}$.
$n_{\chi}$ is the number of electron-hole pairs contained in the states $\vert \chi\rangle\in\{\vert B\rangle, \vert X_{H/V}\rangle\}$.
Finally, the Jaynes-Cummings type coupling of the cavity modes to the QD with coupling constant $g$ is given by:
\begin{align}
  \hat{\mathcal{X}} =-g&\left(\vert G\rangle\langle X_H\vert\hat{a}_H^\dagger+\vert X_H\rangle\langle B\vert\hat{a}_H^\dagger\right.
\notag\\&\left.\hspace{0.25cm}+\vert G\rangle\langle X_V\vert\hat{a}_V^\dagger-\vert X_V\rangle\langle B\vert\hat{a}_V^\dagger\right)+H.c.\,,
\end{align}
where $H.c.$ stands for the Hermitian conjugate and $\vert G\rangle$ is the QD ground state, the energy of which is taken as the zero of energy.
In addition, we account for cavity losses with a rate $\kappa$ by the Lindblad operator:
\begin{equation}
\label{eq:L_cav}
\mathcal{L}_\text{cav}\left[\hat{\rho}\right]=\sum_{\ell=H,V}\frac{\kappa}{2}\left(2\,\hat{a}_\ell\hat{\rho}\hat{a}_\ell^\dagger
-\hat{\rho}\hat{a}_\ell^\dagger\hat{a}_\ell-\hat{a}_\ell^\dagger
\hat{a}_\ell\hat{\rho}\right).
\end{equation}
We assume that the system is initially prepared in the biexciton state, without photons and that the phonons are initially in equilibrium at a temperature $T$.
This can be achieved, e.g., by using two-photon resonant or near-resonant excitation with short coherent pulses \cite{hanschke2018,entangled-photon1,PI_phonon-assisted_biexc_prep-exp,huber2017,reindl2017}, which introduces much less decoherence and time jitter than, e.g., pumping the wetting layer and subsequent relaxation to the biexciton.
The dynamics of the reduced density matrix $\hat{\rho}$ is determined by the equation:
\begin{equation}
\label{eq:dynamic}
\frac{\mathrm{d}}{\mathrm{d}t}\hat{\rho}=-\frac{i}{\hbar}\left[\hat{H},\hat{\rho}\right]_{-}+
\mathcal{L}_\text{cav}\left[\hat{\rho}\right],
\end{equation}
where $[,]_{-}$ denotes the commutator.
As in Ref.~\cite{Different-Concurrences:18}, we evaluate $\hat{\rho}$ numerically in the subspace
spanned by the five states $\vert B,0,0\rangle$, $\vert X_H,1,0\rangle$, $\vert X_V,0,1\rangle$, $\vert G,2,0\rangle$, and $\vert G,0,2\rangle$, where the numbers $n_{H/V}$ in $\vert \chi,n_{H},n_{V}\rangle$ denote the number of $H/V$ photons.
We use a path-integral approach that does not introduce approximations to the model.
This is made possible by recent methodological advances that allow for a natural inclusion of non-Hamiltonian parts of the dynamics (e.g. represented by Lindblad operators) in the path-integral formalism \cite{PI_nonHamil2016} as well as huge improvements of the performance by iterating instead of the augmented density matrix, introduced in the pioneering work of Makri and Makarov \cite{Makri_Theory,Makri_Numerics}, a partially summed augmented density matrix \cite{PI_cQED}.
We quantify the degree of entanglement by the concurrence, a quantity which has a one-to-one correspondence to the entanglement of formation \cite{Wootters:2001}.
To be precise, we use the concurrence of simultaneously emitted photon pairs
\begin{equation}
\label{eq:concurrence}
C = 2\,\frac{\vert\bar{\rho}_{HV}\vert}{\bar{\rho}_{HH}+\bar{\rho}_{VV}}
\end{equation}
(see the Supplemental Material \cite{supp}\nocite{Wootters:2001,Wootters1998,Pfanner2008,multi-time,QuantumStateTomography,Reimer_tune_EB_2011,Ding_tune_EB_2010,Trotta_tune_EB_2012,Trotta_tune_EB_2013,Troiani2006} for further details) that can be calculated directly from the time-averaged occupations $\bar{\rho}_{HH}$, $\bar{\rho}_{VV}$ and coherence $\bar{\rho}_{HV}$ of the states $\vert HH\rangle$ and $\vert VV\rangle$ \cite{carmele11,Different-Concurrences:18,Seidlemann2019}.
We focus on simultaneously emitted photon pairs since experiments \cite{StevensonPRL2008,Bounouar18} agree with theory \cite{EdV,Different-Concurrences:18} that this case is favorable for the entanglement.

\begin{figure*}
\centering
\includegraphics[width=0.9\textwidth]{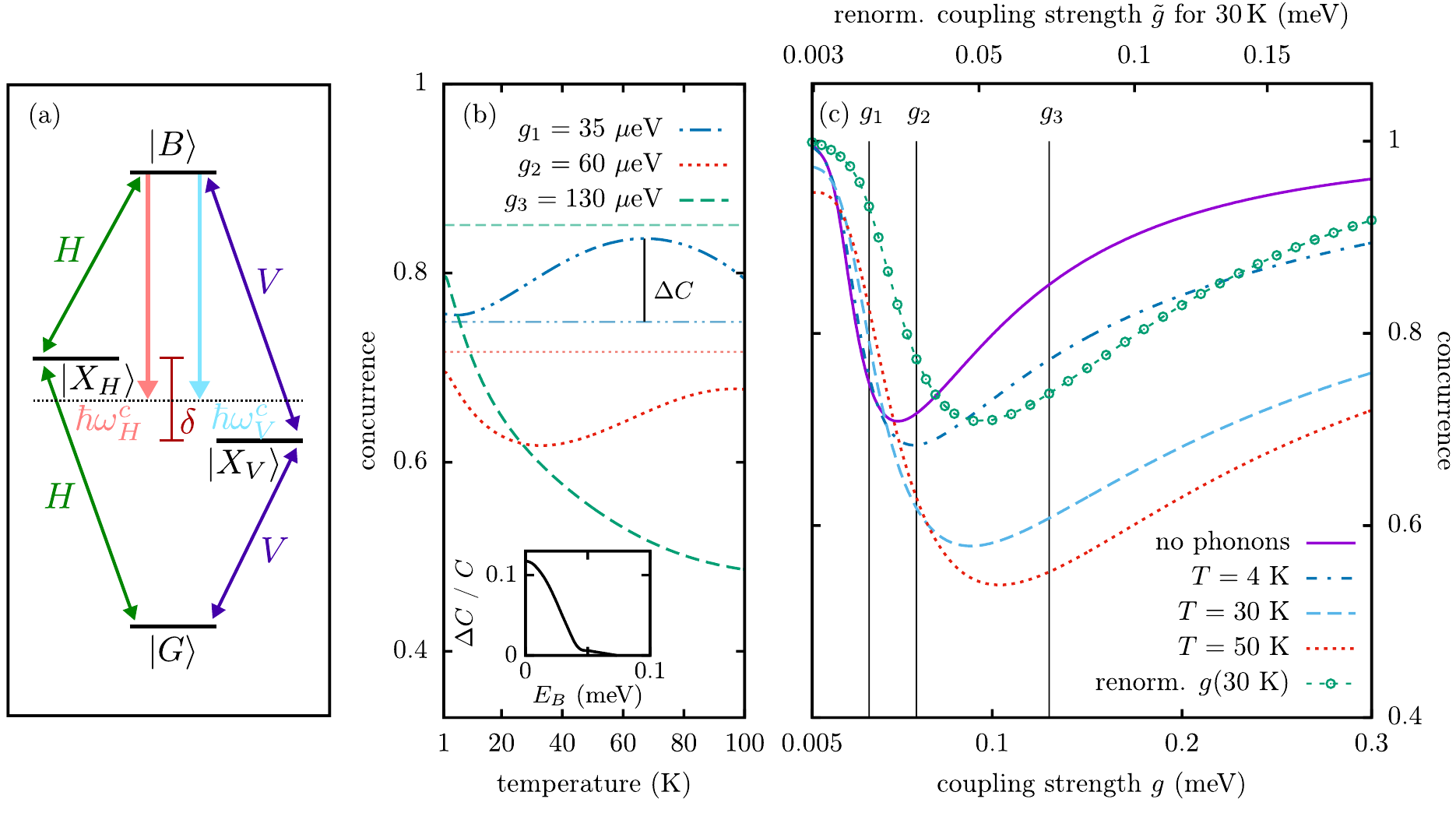}
\caption{(a) Sketch of the level scheme of a QD-cavity  system with finite fine-structure splitting, zero biexciton binding energy and two-photon resonant cavity modes.
(b) Concurrence as a function of the temperature for three selected values of the QD-cavity coupling.
The corresponding values obtained without phonons are drawn as straight (faded) lines with the same linetype.
Inset: difference $\Delta C$ between the maximum concurrence value at finite temperature and the corresponding phonon-free value normalized by the latter as a function of the biexciton binding energy $E_B$ for $g_1 = 35$ \textmu eV.
(c) Concurrence as a function of the QD-cavity coupling for three temperatures together with the phonon-free result. 
In addition $C(\tilde{g}(g))$ is plotted using the phonon-renormalized coupling $\tilde{g}(g)$ for $T=30$ K (indicated on the upper axis), where $C(g)$ is the phonon-free concurrence.
The values of the QD-cavity coupling used in (b) are marked in (c) by vertical lines.
Parameters: $\delta=0.1$ meV, $\kappa=0.025\,\text{ps}^{-1}$, electron (hole) confinement length $a_{e}=3$ nm, $a_{h}=a_{e}/1.15$ where we assume a spherical GaAs-type QD with harmonic confinement. All other parameters, e.g., concerning the phonon coupling are taken from Ref.~\cite{Krummheuer2005}.}
\label{fig1}
\end{figure*} 

\begin{figure*}
\centering
\includegraphics[width=0.9\textwidth]{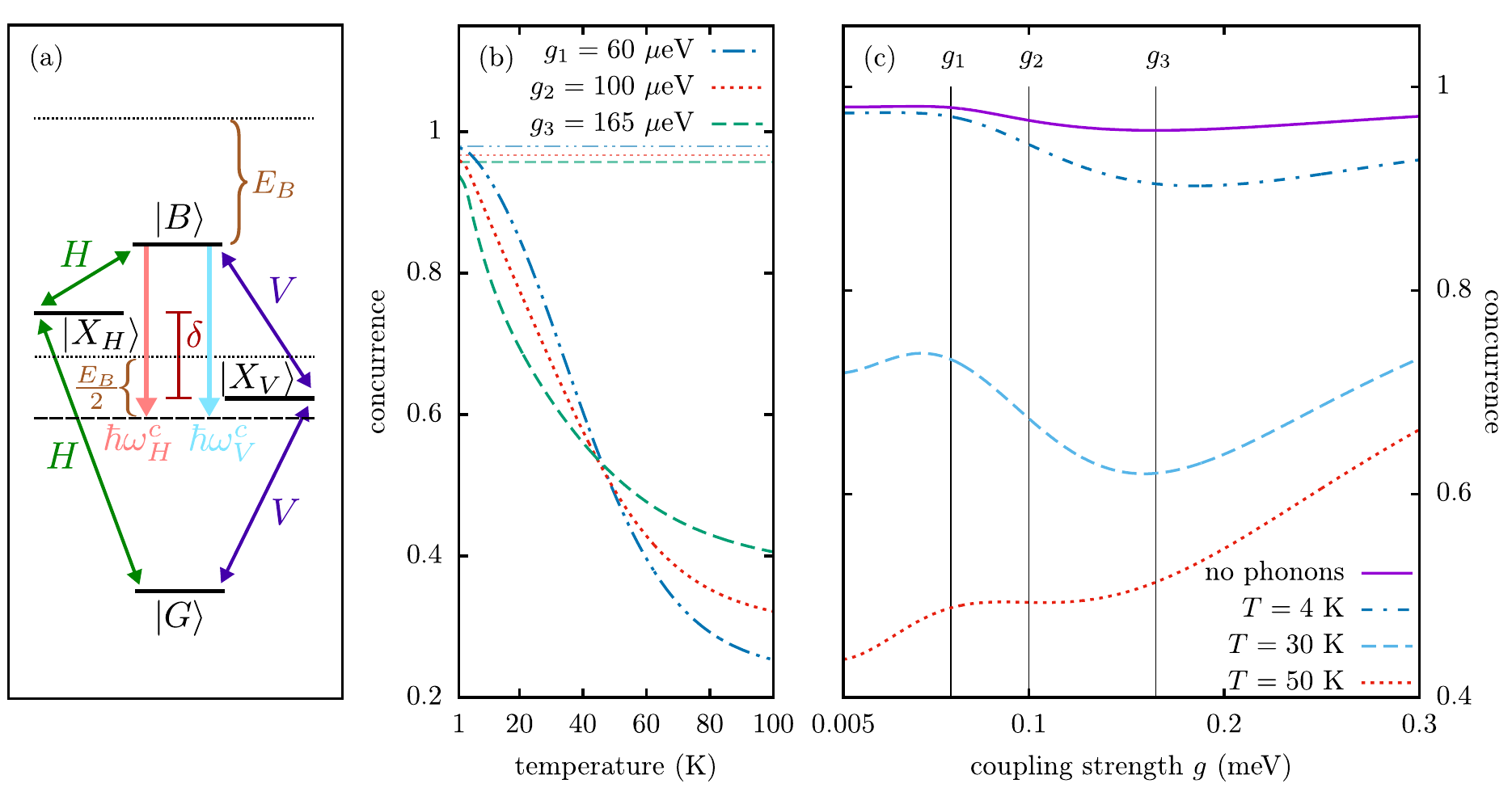}
\caption{(a) Sketch of the level scheme of a QD-cavity system with finite fine-structure splitting, biexciton binding energy $E_{B}=1$ meV and two-photon resonant cavity modes.
(b) Concurrence as a function of the temperature for three selected values of the QD-cavity coupling.
The corresponding values obtained without phonons are drawn as straight (faded) lines with the same linetype.
(c) Concurrence as a function of the QD-cavity coupling for three temperatures together with the phonon-free result.
The values of the QD-cavity coupling used in (b) are marked in (c) by vertical lines.
Apart from $E_{B}$, the same parameters are used as in Fig.~\ref{fig1}.}
\label{fig2}
\end{figure*} 

First, we present results for the situation sketched in Fig.~\ref{fig1}(a) where the excitons have a finite fine-structure splitting $\delta=\hbar(\omega_{H}-\omega_{V})$, the biexciton binding energy is zero and both cavity modes are tuned to the two-photon resonance $2\omega^{c}_{H}=2\omega^{c}_{V}=\omega_{H}+\omega_{V}-\omega_{B}$.
In the situation with phonons, these QD energies denote the polaron-shifted ones. 
To compare QD-cavity systems with identical energy relations, the energy values are kept the same in the corresponding phonon-free calculations thus keeping the polaron shifts.

Figure~\ref{fig1}(b) displays the temperature dependence of the concurrence for three values of the QD-cavity coupling.
Only the result for $g=130\, \mu\text{eV}$ agrees with the common expectation that the entanglement should monotonically decrease with temperature.
In contrast, for $g=60\, \mu\text{eV}$ and $g=35\, \mu\text{eV}$, unusual nonmonotonic $T$ dependences are found.
Most interestingly, for $g=35 \,\mu\text{eV}$, the concurrence is noticeably higher than the corresponding value obtained without phonons in the entire $T$ range that we consider ($T\in[1\,\text{K},100\,\text{K}]$); i.e., for certain values of $g$ we find indeed a phonon-induced enhancement of entanglement while in other cases the expectation that phonons reduce the entanglement is confirmed.

The reason for this remarkable behavior becomes apparent when looking at the $g$ dependence of the concurrence in Fig.~\ref{fig1}(c).
Already without phonons, the concurrence is a nonmonotonic function of $g$ (purple curve) with a pronounced minimum reached roughly for $g\simeq\delta/2$.
Dividing Eq.~(\ref{eq:dynamic}) by the coupling strength $g$ and leaving out the coupling to phonons, the system dynamics is described by the rescaled quantities $t'=gt$, $g'=g/g=1$, $\delta'=\delta/g$, and $\kappa'=\kappa/g$. 
Since the concurrence is the asymptotic value of the normalized coherence at long averaging times \cite{Different-Concurrences:18}, the rescaling of the time is irrelevant.
For large values of $g$, both parameters $\delta'$ and $\kappa'$ tend to zero.
This implies that the concurrence approaches unity for large coupling strengths because the which-path information disappears for a vanishing splitting and thus the concurrence is one \cite{carmele11,Different-Concurrences:18}. 
For very small QD-cavity couplings, $\kappa'$ and $\delta'$ become arbitrarily large.
Therefore, the sequential single photon decay via the intermediate exciton states becomes strongly off-resonant and is thus negligible compared with contributions from a direct two-photon transition which is always resonant in the present case \cite{Seidlemann2019}.
Since the which-path information is contained only in the sequential decay the concurrence approaches unity again.
But for finite splittings, the concurrence is smaller than one and thus a minimum must appear at a certain coupling strength $g$.

When phonons are accounted for, the minimum is lowered and shifted to a higher coupling strength depending on the temperature.
We attribute the shift to the well known effect of phonon-induced renormalization of the light-matter coupling \cite{ramsay:10a}.
To support this assignment we have estimated the renormalized coupling $\tilde{g}(g)$  as in Ref.~\cite{hybrid} by fitting equations with phenomenological renormalizations of a resonantly driven two-level system to path-integral calculations.
The results are shown in the Supplemental Material \cite{supp}.
If the only effect introduced by phonons was the $g$ renormalization, then the value of the concurrence found without phonons at a particular value of $g$ should be shifted by phonons to $\tilde{g}(g)$.
Indeed, in Fig.~\ref{fig1}(c) we have plotted $C(\tilde{g}(g))$ using the phonon-renormalized coupling $\tilde{g}(g)$ for $T=30$ K, where $C(g)$ is the concurrence in the phonon-free case (green curve with circles).
We find that, despite the crudeness of the estimation for $\tilde{g}(g)$, the minimum of the shifted curve agrees even quantitatively well with the minimum found in the full path-integral simulation for this temperature (red dotted curve).
Since the shift is larger for higher temperatures, displacing the phonon-free curve necessarily leads to higher values of the shifted curves in regions where the phonon-free concurrence is monotonically decreasing with $g$.  
Consequently, in this region, phonon-induced enhancement appears for a finite $g$ range.

The total effect of phonons is, however, not merely a shift but also a lowering of the curves with rising temperature, which is indeed due to the dephasing action of phonons.
It is important for obtaining a phonon-induced entanglement that the gain in entanglement resulting from the shift of the phonon-free curve due the phonon-induced $g$ renormalization is not destroyed by the overall lowering of the concurrence caused by the decoherence.
Figure~\ref{fig1}(c) demonstrates that it is indeed possible that the renormalization-induced shift overcompensates the dephasing action.
Additionally, when accounting for pure dephasing by introducing a phenomenological rate \cite{Jahnke2012}, the  phonon-induced enhancement disappears (see the Supplemental Material \cite{supp}).
This result reaffirms the $g$ renormalization as the main origin of the effect, since it is absent in the phenomenological model.

It is instructive to contrast the above findings with simulations for the more commonly considered situation sketched in Fig.~\ref{fig2}(a), where the biexciton binding energy has the finite value $E_{B}=1$ meV and the cavity modes are in resonance with the two-photon transition to the biexciton.
Again, the phonon-free curve exhibits a minimum which is, however, rather flat [purple line in Fig.~\ref{fig2}(c)].
In the limit $g\to\infty$ the concurrence approaches unity since the argument given for the case of vanishing biexciton binding energy applies here as well.
For the case that both $g/(\frac{1}{2}E_{B})$ and $\delta/E_{B}$ are small parameters, it has been shown analytically in Ref.~\cite{Different-Concurrences:18} that the phonon-free concurrence approaches
$\left[\left(E_{B}^{2}-\delta^{2}\right)/\left(E_{B}^{2}+\delta^{2}\right)\right]$, which is smaller than one for a finite $\delta$.
Including phonons, the reduction of the concurrence for small $g$ values is strongly magnified as seen in Fig.~\ref{fig2}(c).
Overall, the dephasing action induced by phonons is so strong that the line shape of the concurrence as a function of $g$ is significantly deformed, and the effects related to a renormalization of $g$ cannot be identified.
As a consequence, the concurrence monotonically decreases with rising temperature and always stays below the phonon-free calculation for all values of $g$ as exemplarily shown in Fig.~\ref{fig2}(b).
This demonstrates that the phonon-induced enhancement of entanglement described above can only occur when the $g$-renormalization effects dominate over the phonon-induced dephasing.
The stronger phonon-induced dephasing for $E_{B}$ on the order of a few meV compared with vanishing $E_{B}$ has been explained recently \cite{Seidlemann2019} by noting that the energies bridged by phonon-assisted processes are closer to the maximum of the phonon spectral density in the former case.

It is worthwhile to note that phonon-induced enhancement of photon entanglement is not restricted to the singular case of vanishing $E_B$ but rather appears for a finite range of binding energies as demonstrated in the inset of
Fig.~\ref{fig1}(b).
The difference $\Delta C$ between the maximum concurrence value at finite temperatures and the corresponding phonon-free value is positive clearly for an extended range.
Further analysis (shown in the Supplemental Material \cite{supp}) reveals that the effect can be observed as long as $E_B\lesssim\delta/2$ holds for our realistic parameters.

We note in passing that the situation considered in Fig.~\ref{fig1} is not the only one where the conditions for phonon-induced entanglement are realized.
This phenomenon can also be observed in a system with weak biexciton binding and degenerate excitons where which-path information is introduced by a finite splitting of the cavity modes (see the Supplemental Material \cite{supp}).
There the concurrence calculated without phonons is again a nonmonotonic function of $g$, which exhibits even more than one extremum.
Also in this case, the phonon-induced renormalization is strong enough to evoke a phonon-induced entanglement for finite parameter ranges.

In conclusion, we demonstrate that  phonon-induced renormalizations of the dot-cavity coupling can overcompensate decoherence effects and shift the system to a region of higher entanglement.
In combination with a nonmonotonic dependence of the phonon-free concurrence, this can result in  a nonmonotonic temperature dependence of the concurrence. Most interestingly, the concurrence can even reach values above the phonon-free level, thus causing phonon-induced photon entanglement.

\acknowledgments
M.~Cygorek thanks the Alexander-von-Humboldt foundation for support through a Feodor Lynen fellowship. 
A.~Vagov acknowledges the support from the Russian Science Foundation under the Project No. 18-12-00429, which was used to study dynamical processes nonlocal in time by the path-integral approach.
This work was also funded by the Deutsche Forschungsgemeinschaft (DFG, German Research Foundation) - project Nr. 419036043.

\bibliography{PIbib}
\end{document}